\documentclass[twocolumn,showpacs,aps,prd,superscriptaddress]{revtex4}

\usepackage{graphicx}
\usepackage{dcolumn}
\usepackage{amsmath}
\usepackage{epsfig}
\usepackage{dcolumn}

\newcolumntype{.}{D{.}{.}{-1}}
\newcolumntype{,}[1]{D{.}{.}{#1}}
\newcolumntype{p}{D{\%}{\%}{3}}
\newcolumntype{a}{D{.}{\to}{-1}}

\input babarsym.tex

\def\mes        {\mbox{$m_{\rm ES}$}\xspace}

\newcommand{\BABARPubYear}    {06}
\newcommand{\BABARPubNumber}  {021}
\newcommand{\SLACPubNumber} {11801}

\def\figurebox#1#2#3{%
    \def\arg{#3}%
    \ifx\arg\empty
    {\hfill\vbox{\hsize#2\hrule\hbox to #2{\vrule\hfill\vbox to #1{\hsize#2\vfill}\vrule}\hrule}\hfill}%
    \else
    {\hfill\epsfbox{#3}\hfill}%
    \fi}

\newcommand {\vecp} {\ensuremath{\kern 0.2em\vec{\kern 0.1em p}{}\xspace}}

\newcommand{\BtoDstOmgPi}{\Bzb \to \Dstarp \omega \pim}
\newcommand{\BtoDstFourPi}{\Bzb \to \Dstarp \pim \pip \pim \piz}

\begin{document}

\begin{flushleft}
\babar-PUB-\BABARPubYear/\BABARPubNumber\\
SLAC-PUB-\SLACPubNumber\\
\end{flushleft}

\title{
{\large \bf
Study of the decay {\boldmath $\BtoDstOmgPi$}}}

%
\author{B.~Aubert}
\author{R.~Barate}
\author{M.~Bona}
\author{D.~Boutigny}
\author{F.~Couderc}
\author{Y.~Karyotakis}
\author{J.~P.~Lees}
\author{V.~Poireau}
\author{V.~Tisserand}
\author{A.~Zghiche}
\affiliation{Laboratoire de Physique des Particules, F-74941 Annecy-le-Vieux, France }
% ...
\collaboration{The \babar\ Collaboration}
\noaffiliation

\begin{abstract}
We report on a study of the decay $\Bzb \to \Dstarp \omega \pi^-$
with the \babar\ detector at the PEP-II $B$-factory at the 
Stanford Linear Accelerator Center.
Based on a sample of 232 million \BB decays, we measure
the branching fraction ${\cal B}(\Bzb \to \Dstarp \omega \pi^-) = 
(2.88 \pm 0.21{\rm(stat.)} \pm 0.31{\rm(syst.)}) \times 10^{-3}$.  
We study the invariant mass spectrum of the
$\omega \pi^-$ system in this decay.
This spectrum is in good agreement with expectations
based on factorization and the measured spectrum in $\tau^- \to \omega \pi^- 
\nut$.    We also measure the polarization of the \Dstarp as a
function of the $\omega \pi^-$  mass.  In the mass region 1.1 to 1.9 
GeV we measure the fraction of longitudinal polarization of
the $\Dstarp$ to be $\Gamma_L/\Gamma = 0.654 \pm 0.042{\rm(stat.)} \pm 0.016{\rm(syst.)}$.  This 
is in agreement with the expectations from heavy-quark effective theory and
factorization assuming that the decay proceeds as
$\Bzb \to \Dstarp \rho(1450)$, $\rho(1450) \to \omega \pi^-$.
\end{abstract}


\maketitle

\section{Introduction}
Factorization is a powerful tool
to describe hadronic decays of the $B$-meson.  
According to factorization, the matrix element of four-quark
operators can be written as the product of matrix elements of
two two-quark operators~\cite{FactIntro}.
Thus, the process
$b \to c W^{*},~W^{*} \to q\bar{q}'$ 
(where $q = d$ or $s$, $q' = u$ or $c$)
 can be ``broken up'' into
two pieces, the $b \to c$ transition and the hadronization from 
$W^* \to q\bar{q}'$ decay.

\begin{figure} [htb]
\centerline{
\epsfig{file=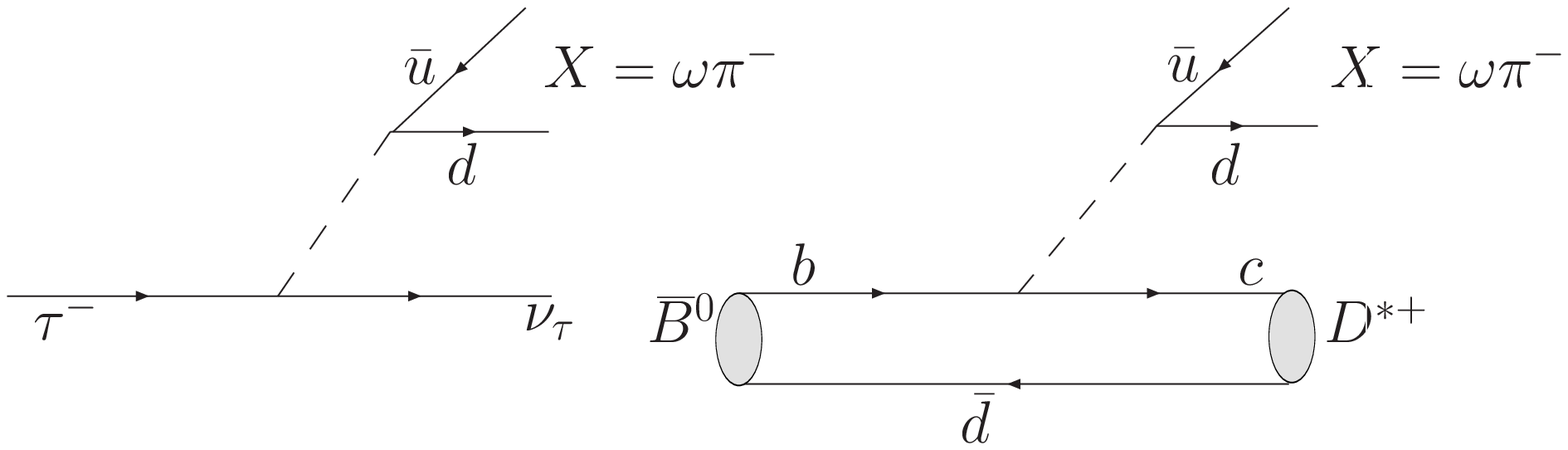,width=0.95\linewidth}}
\caption{Feynman diagrams for
$\tau \to \omega \pi \nu$ and $\BtoDstOmgPi$.}
\label{fig:bomegapi_feyn}
\end{figure}

Ligeti, Luke, and Wise have
proposed an elegant test of factorization~\cite{LLW}.
In this test, data from $\tau \to X \nu$, 
where $X$ is a hadronic system, is used 
to predict the properties of $B \to D^{*} X$ (see
Fig.~\ref{fig:bomegapi_feyn}).
If $X$ is composed of two or more particles not dominated 
by a single narrow resonance, factorization can be tested 
in different kinematic regions.

In the event that $X$ is a multi-body system, it is possible 
that some fraction of the 
hadronic system could be emitted in association with 
the $B \to D^{(*)}$ transition instead of the 
hadronization from $W^* \to q\bar{q}'$ decay.
In the case of $X \equiv \omega \pi^-$, the pion
must come from the $W^*$ to conserve charge.  
It is unlikely that the
$\omega$ could be produced from the lower vertex in
Fig.~\ref{fig:bomegapi_feyn}~\cite{LLW,CDstOmg}.  
Furthermore the $\omega \pi^-$ state is not associated
with any narrow resonance, so that a wide range in
$\omega \pi^-$ invariant mass can be studied.
As the branching fraction for $\BtoDstOmgPi$ is large 
($\approx$ 0.3\%), this decay provides 
a good laboratory for the study of factorization.
                                       
The branching fraction for $\BtoDstOmgPi$ 
has been measured 
by the CLEO collaboration, using a sample of 9.7 
million $B\bar{B}$ pairs collected at the $\FourS$ resonance, to be
$(2.9 \pm 0.3{\rm (stat.)} \pm 0.4{\rm (syst.)}) \times 10^{-3}$~\cite{CLEO}.
They also extracted the spectrum of
$m_X^2$, the square of the invariant mass of the $\omega \pi$ system.
This spectrum is found to be in agreement with theoretical
expectations~\cite{LLW}.  In addition, the CLEO collaboration
studied the related decay
$B \to D \omega \pi$ and concluded that this decay is dominated
by the broad $\rho(1450)$ intermediate resonance; {\it i.e.}, 
$B \to D \rho(1450)$, $\rho(1450) \to \omega \pi$.
Assuming that this intermediate state also dominates in $\BtoDstOmgPi$, 
factorization can be used to predict the 
polarization of the $D^*$ with the aid of 
heavy-quark effective theory (HQET)
and data from semileptonic $B$-decays~\cite{neubert}.
These predictions are in agreement with the CLEO 
result for the longitudinal polarization fraction, 
$\Gamma_L/\Gamma = (63 \pm 9)$\%~\cite{CLEO}.

In this paper we study the decay $\BtoDstOmgPi$
with a larger sample of $B$ decays 
than available in the original CLEO study.
We present measurements of the
branching fraction, the Dalitz plot distribution, the 
$m_X^2$ spectrum, the $m_{D^* \pi}$ distribution, 
and the $D^*$ polarization as a function of $m_X$.

\section{The \babar\ Dataset and Detector}
The results presented in this paper are based on $232 \times 10^6$ 
$\FourS \to B\Bbar$ decays, corresponding to an integrated luminosity
of 211 fb$^{-1}$.  The data were collected between 1999 and 2004 with the
\babar\ detector~\cite{babar} at the \pep2\ \BF\ at SLAC.
In addition a 22 fb$^{-1}$ off-resonance data sample, with
center-of-mass energy 40 MeV below the \FourS resonance, is used to
study backgrounds from continuum events, $e^+e^- \to q\bar{q}$ 
($q = u,d,s,$ or $c$).

Charged-particle tracking is provided by a five-layer 
double-sided silicon vertex tracker (SVT) and a 40-layer drift chamber (DCH), 
operating within a 1.5-T magnetic field.  Energy depositions are 
measured with a CsI(Tl) electromagnetic calorimeter (EMC).  
Charged particles are identified from ionization energy loss 
(d$E$/d$x$) measurements in the SVT and DCH, and from 
the observed pattern of Cherenkov light in an internally reflecting 
ring imaging detector.

\section{Analysis Strategy}
\label{sec:strategy}
Starting from the set of
reconstructed charged 
tracks and energy deposits within the EMC, we
select events that are kinematically consistent with
$\BtoDstOmgPi$ in the following decay 
modes: $D^{*+} \to D^0 \pi^+$, with $D^0 \to K^- \pi^+$, 
$K^- \pi^+ \pi^+ \pi^-$, or $K^- \pi^+ \pi^0$, and 
$\omega \to \pi^+ \pi^- \pi^0$. 
Charge-conjugate modes are implied throughout this paper.

In the reconstruction chain, the invariant mass requirement 
on the $\pi^+ \pi^- \pi^0$ system that forms the $\omega$ candidate
is kept loose.  We then select 
``signal'' or ``sideband'' candidates 
depending on whether the reconstructed
$\pi^+ \pi^- \pi^0$ mass is consistent with the $\omega$ hypothesis.
Kinematic distributions of interest, such as the $m_X^2$ spectrum, 
are obtained
by subtracting, with appropriate weights, the distributions for
signal and sideband events.  This subtraction accounts
for all sources of backgrounds, including backgrounds from 
$\BtoDstFourPi$, 
on a statistical basis.  
This is because, as we will demonstrate in Section~\ref{sec:bg},
background sources with real $\omega$ decays are negligible.

The event reconstruction efficiency is determined from 
simulated Monte Carlo events, where the response of the 
\babar\ detector is modeled using the GEANT4~\cite{GEANT} program. 
Efficiency-corrected
kinematic distributions are obtained by assigning a weight to each 
event.  This weight is equal to the inverse of the efficiency to 
reconstruct that particular event given its kinematic properties.  
This procedure,
which is independent of assumptions on the dynamics of the
$\BtoDstOmgPi$ decay,
is discussed in Section~\ref{sec:eff}.

\section{Event Selection Criteria}
The event selection criteria are optimized based on studies of 
off-resonance data, and simulated
$B\bar{B}$ and continuum events.

Photon candidates are constructed from calorimeter clusters
with lateral profiles consistent with photon showers and with 
energies above 30 MeV.  
Neutral pion candidates are formed from
pairs of photon candidates with invariant mass between
115 and 150 MeV and energy above 200 MeV, where the $\pi^0$
mass resolution is 6.5 MeV.  In order to improve 
resolution, $\pi^0 \to \gamma \gamma$ candidates are
constrained to the world average $\pi^0$ mass~\cite{PDG}.

The kaon-candidate track used to reconstruct the 
$D^0$ meson must satisfy a set of
kaon identification criteria. 
The kaon identification efficiency
depends on
momentum and polar angle, and is typically about 93\%.
These requirements provide a rejection factor of order 10
against pions.
For each $D^0 \to K^- \pi^+ \pi^0$ candidate, we calculate the
square of the decay amplitude ($|A|^2$) based on
the kinematics of the decay products and the
known
properties of the Dalitz plot for this decay~\cite{dalitz}.
We retain candidates
if $|A|^2$ is greater than 2\% of its
maximum possible value.
The efficiency of this requirement is 91\%.
Finally, the measured invariant mass of $D^0$ candidates 
must be within 15 MeV of the world average 
$D^0$ mass~\cite{PDG} 
for $D^0 \to K^- \pi^+$ and $D^0 \to K^- \pi^+ \pi^- \pi^-$,
and 25 MeV for $D^0 \to K^- \pi^+ \pi^0$.  The experimental 
resolution is about 6 MeV for $D^0 \to K^- \pi^+$, 
$K^- \pi^+ \pi^- \pi^-$, and 10 MeV for $D^0 \to K^- \pi^+ \pi^0$.

We select $D^{*+}$ candidates by combining $D^0$ candidates 
with an additional track, assumed to correspond to a pion.
We require the measured 
mass difference $\Delta m \equiv m(D^{*+})-m(D^0)$
to be between 143.4 and 147.4 MeV.  The resolution on this
quantity is 0.3 MeV with non-Gaussian behavior 
due to the reconstruction of the low momentum pion 
from $D^*$ decay.

In the rest frame of the $\Bzb$, as $m_X^2$ increases 
the $D^{*+}$ is produced with decreasing energy.  At high $m_X^2$, 
or equivalently low $D^{*+}$ energy, the reconstruction 
efficiency drops as $\cos \theta_D \to 1$, where $\theta_D$ is the 
angle 
between 
the daughter $D^0$ and the direction opposite the flight of the $\Bzb$ 
in the $D^{*+}$ rest frame.
We exclude the region of low acceptance 
($\cos \theta_D > 0.8$ for $8 \leq m_X^2 < 9$ GeV$^2$, 
$\cos \theta_D > 0.6$ for $9 \leq m_X^2 < 10$ GeV$^2$, and
$\cos \theta_D > 0.4$ for $m_X^2 \geq 10$ GeV$^2$)
from our event selection.  The 
effect on the final results is very small, as will be discussed in 
Section~\ref{sec:results}.

We form $\omega$ candidates from a pair of oppositely-charged
tracks, assumed to be a $\pi^+\pi^-$ pair, and a $\pi^0$ candidate.
In order to 
keep signal and sideband candidates (see Section~\ref{sec:strategy})
we impose only the very loose requirement that the invariant
mass of the $\omega$ candidate be within 70 MeV of the 
world average $\omega$ mass~\cite{PDG}.
(The natural width of the $\omega$ resonance is 8.5 MeV and
the experimental resolution is 5.6 MeV.)

In order to reduce combinatoric
backgrounds, we impose a requirement on the kinematics
of the $\omega$ decay~\cite{Perkins}.  
This is done by first defining two Dalitz plot coordinates:
$X \equiv 3T_0/Q - 1$ and 
$Y \equiv \sqrt{3}(T_{+} - T_{-})/Q$, where
$T_{\pm,0}$ are the kinetic energies of the pions
in the $\omega$ rest frame and $Q \equiv T_{+} + T_{-} + T_0$.
Next, we define the normalized square of the distance from the
center of the Dalitz plot, $R^2 \equiv (X^2 + Y^2) / (X_b^2 + Y_b^2)$,
where $X_b$ and $Y_b$ are the coordinates of the intersection between the
kinematic 
boundary of the Dalitz plot and a line passing through $(0,0)$ 
and $(X,Y)$.  Since the Dalitz plot density for real $\omega$
decays peaks at $R=0$, we impose the requirement $R<0.85$.
This requirement is 93\% efficient for signal 
and rejects 25\% of the combinatorial background.

We reconstruct a $B$-meson candidate by combining a
$D^{*+}$ candidate, an $\omega$ candidate, and an additional
negatively charged track.  A $B$-candidate is characterized
kinematically by the energy-substituted mass 
$\mes \equiv \sqrt{(\frac{1}{2} s + \vec{p}_0\cdot \vec{p}_B)^2/E_0^2 - p_B^2}$, 
where $E$ and $p$ denote energy and momentum measured in the lab frame, 
the subscripts $0$ and $B$ refer to the
initial \FourS and $B$ candidate, respectively, and 
$s$ represents the square of the energy of the $e^+e^-$ 
center of mass (CM) system.
For signal events we expect $\mes \approx M_B$
within the experimental resolution of
about 3 MeV, where $M_B$ is the world average $B$ mass~\cite{PDG}.
In the same fashion, the energy difference 
$\Delta E \equiv E_B^*-\frac{1}{2}\sqrt{s}$,
where the asterisk denotes the CM frame, is expected to 
be nearly zero for signal $B$ decays.

The $\Delta E$ resolution is approximately 25 MeV
in the $K^- \pi^+ \pi^0$ mode and 20 MeV in the
other modes, with non-Gaussian tails towards
negative values due to energy leakage in the EMC.
We select $B$ candidates with 
a $D^0 \to K^- \pi^+ \pi^0$ as long as 
$-70 \leq \Delta E \leq 40$ MeV, and we require 
$-50 \leq \Delta E \leq 35$ MeV for the other modes.  

In order to further reduce the number of events from 
continuum backgrounds 
we make two additional requirements.   
First, we require $|\cos\theta_B| < 0.9$, where 
$\theta_B$ is the decay angle of the $B$ candidate with 
respect to the $e^-$ beam direction in the CM frame.  
For real $B$ candidates, $\cos\theta_B$ follows a 
$1 - x^2$
distribution, while the distribution is essentially flat for 
$B$ candidates formed from random combinations of tracks. 
Second, we impose a requirement on a Fisher discriminant~\cite{Fisher}
designed to differentiate between spherical $B\bar{B}$ events
and jet-like continuum events.
This discriminant is constructed  
from the quantities $L_0 = \sum_i{p^*_i}$ and $L_2 =
\sum_i{p^*_i \cos^2\alpha^*_i}$. Here, $p^*_i$ is 
the magnitude of the momentum and
$\alpha^*_i$ is the angle with respect to the thrust axis of the $B$
candidate of tracks and clusters not used to reconstruct the $B$,
all in the CM frame.  The requirements on $|\cos\theta_B|$ and 
the Fisher discriminant are 95\% efficient for signal and 
reject nearly 40\% of the continuum background.

The reconstruction of the $\BtoDstOmgPi$ decay is 
improved by refitting the momenta of the decay products of the 
$\Bzb$, taking into account kinematic and geometric 
constraints.  The kinematic constraints are based
on the fact that their decay products must originate
from a common point in space.  The entire decay chain is fit 
simultaneously in order to account for any correlations between 
intermediate particles. 

If more than one $B$ candidate is found in a given event 
with $\mes$ $>$ 5.2 GeV, and passes selection requirements, 
we retain the best candidate based on a $\chi^2$ algorithm 
that uses the measured values, world average values, and resolutions 
of the $D^0$ mass and the mass difference $\Delta m$.  
We omit the $\omega$ candidate mass information from arbitration in order 
to avoid introducing a bias in the $\omega$ mass distribution, 
since this distribution is used extensively throughout the analysis.

\section{Event yield}
\label{sec:yield}

\begin{figure}[hbt]
\begin{center}
\epsfig{file=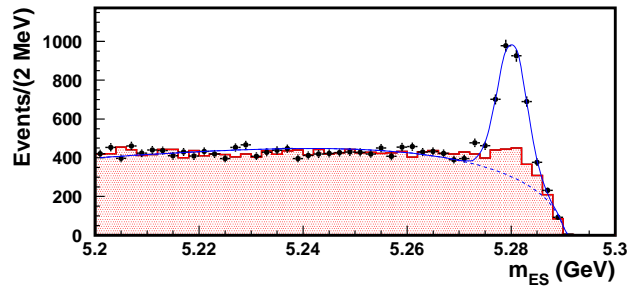,width=\linewidth}
\caption{\protect $\mes$ distributions for candidates with
reconstructed $\omega$ mass
in the signal (points) and sideband (shaded histogram) regions.
The distribution for events in the sideband region has been
rescaled to match the expected background in the $m_{\omega}$ 
signal region.
The fitted function is described in the text.}
\label{fig:mes}
\end{center}
\end{figure}

In Fig.~\ref{fig:mes} we show
the $\mes$ distribution for candidates with reconstructed 
$\pi^+ \pi^- \pi^0$ mass ($m_{\omega}$)
in the signal and sideband regions, which are defined
as $|m_{\omega} - m_{\omega}^{PDG}| < 20$ MeV and
$35 < |m_{\omega} - m_{\omega}^{PDG}| < 70$ MeV,
respectively, where $m_{\omega}^{PDG}$ is the 
world average $\omega$ mass~\cite{PDG}.

The $\mes$ distribution for the
$m_{\omega}$ signal region has been fitted to the sum of a threshold
background function~\cite{ARGUS} and a 
Gaussian distribution centered at $M_B$.  The distribution
for the $m_{\omega}$ sideband region demonstrates the 
presence of a background component, which peaks in $\mes$ 
but not in $m_{\omega}$, that is not well described by the
threshold function.
Monte Carlo studies indicate
that approximately one-third of this component is
due to signal events where the $\omega$ is mis-reconstructed.
These are, for example, events where one of the pion tracks in the
$\omega$ decay is lost and is replaced by a track from the decay 
of the other $B$ in the event.  The remaining two-thirds of the 
$\mes$ peaking background component is due to 
$\Bzb \to D^{*+} \pi^+ \pi^- \pi^0 \pi^-$ events.

We extract the event yield from a binned $\chi^2$ fit of 
the $m_{\omega}$ distribution for events with $\mes$ $>$ 5.27 GeV.
The data distribution is modelled as 
the sum of a Voigtian function and a linear background function.
(The Voigtian is the convolution of a Breit-Wigner with a Gaussian 
resolution function.)  The width of the Breit-Wigner is fixed 
at 8.5 MeV, the world average width of the $\omega$.  The mass
of the $\omega$, the Gaussian resolution term, and the parameters
of the linear function are free in the fit.

\begin{figure}[hbt]
\begin{center}
\epsfig{file=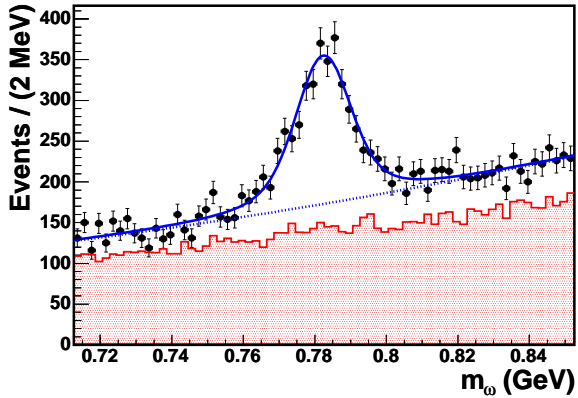,width=\linewidth}
\caption{\protect Distribution of reconstructed 
$m_{\omega}$ for events with $\mes > 5.27$ GeV (points) and 
events with  $5.20 < \mes < 5.25$ GeV (shaded histogram).  
The superimposed fit is described in the text.  The events 
from the $\mes$ sideband have been scaled to the expected 
background from an $\mes$ fit to events with 
$|m_{\omega} - m_{\omega}^{PDG}| < 70$ MeV 
({\it i.e.}, the range shown in this figure).}
\label{fig:momg-yield}
\end{center}
\end{figure}

The $m_{\omega}$ distribution and the associated fit is 
shown in Fig.~\ref{fig:momg-yield}.  The yield,
defined as the number of events in the Voigtian
with $|m_{\omega} - m_{\omega}^{PDG}| < 20$ MeV,
is $1799 \pm 87$ events.  
The Gaussian resolution returned by the fit 
as well as the mean of the Breit-Wigner
are consistent with the value we find in Monte Carlo simulations of 
$\BtoDstOmgPi$ events.  
In Fig.~\ref{fig:momg-yield} we also 
include the $m_{\omega}$ distribution for events 
with $5.20 < \mes\ < 5.25$ GeV (the $\mes$ sideband).  
This background distribution 
has been scaled to the number of background events expected 
from a fit to the $\mes$ distribution where we require 
$|m_{\omega} - m_{\omega}^{PDG}| < 70$ MeV.  The difference 
between the number of observed events away from the $m_{\omega}$ 
peak and the number of background events predicted from the 
$\mes$ sideband is due to the background component that peaks in 
$\mes$.

The validity of the yield extraction relies 
on the assumption that the background is linear
in $m_{\omega}$, and, most importantly, that 
there are no sources of combinatoric backgrounds
that include real $\omega$ decays.  
The results shown in Fig.~\ref{fig:momg-yield}
imply that there is no significant component of real 
$\omega$ decays in the background.
To verify this, we have examined and fit the $m_{\omega}$
distribution for data events in the $\mes$ sideband as well 
as the distribution for Monte Carlo simulations of 
$B\bar{B}$ events, excluding $\BtoDstOmgPi$.
We find that the distributions are well modelled by linear functions.
There is no evidence of a real $\omega$ component in the background. 
We estimate that this component can affect the yield extraction of
Fig.~\ref{fig:momg-yield} at most at the few percent level.

We also divide our dataset into three independent sub-datasets,
according to the three $D^0$ decay modes that we consider.  The
fits to these sub-datasets yield consistent results.

\section{Background Subtraction}
\label{sec:bg}

In this work we are interested in studying a number of
kinematic distributions for $\BtoDstOmgPi$,
such as the $m^2_{X}$ distribution, where $m_{X}$ is the
invariant mass of the $\omega \pi$ system.  The measurements
of these distributions need to account for the presence of 
background in the sample and for the fact that the 
signal reconstruction
efficiency is not constant over the Dalitz plot.

We use distributions for $\omega$ sideband events
to remove the effects of the background in the $\omega$ 
signal region on a statistical basis, and we use Monte Carlo 
simulations to correct for efficiency effects.
This is accomplished as follows:

\begin{enumerate}

\item The simulation of 
$\BtoDstOmgPi$ events is used to calculate the 
signal reconstruction efficiency $\epsilon(\vec{x})$, where
$\vec{x}$ is the set of quantities that specify the kinematics
of a given event.  The procedure used to determine
$\epsilon(\vec{x})$ is discussed in Section~\ref{sec:eff}.

\item In the absence of background, we would calculate 
the number of events
corrected for efficiency in a given bin of $m^2_X$ as
 
\begin{equation}
N(m^2_X) = \sum_{\rm signal} \frac{1}{\epsilon(\vec{x}_i)},
\label{eq:bgsub}
\end{equation}

\noindent where the sum is over signal events in a given $m^2_X$ bin and
$\vec{x}_i$ is the set of kinematic quantities for the $i$-th event
in the sum.

\item As mentioned above, the background subtraction is
performed using the $m_{\omega}$ 
sideband.  Thus, Eq.~\ref{eq:bgsub} is modified to be 

\begin{equation}
N(m^2_X) = \sum_{\rm signal} \frac{1}{\epsilon(\vec{x}_i)}~~-~~
\frac{4}{7} \beta \sum_{\rm sideband} \frac{1}{\epsilon(\vec{x}_j)}
\label{eq:bgsub-effcorr}
\end{equation}
\noindent where the first sum is just as before, while the second 
sum is over $\omega$-mass sideband events in the given bin of $m_X^2$
and $\vec{x}_j$ represents the set of kinematic quantities 
for the $j$-th event in the sideband event sample.
The same efficiency is used for both the signal and sideband 
event samples.
The factor of $\frac{4}{7}$ is needed to adjust for 
the relative size of the $\omega$ signal 
and sideband 
regions.  The additional factor of $\beta$ is ideally equal to one, and it
is introduced to correct for any possible bias in the background subtraction
procedure, as will be discussed below.
\end{enumerate}

The allowed kinematic limits for some
variables, such as $m^2_X$, 
are not the same for $\omega$ signal and sideband events.  
Therefore, the values of these 
variables for events in the $\omega$ sideband region
are linearly rescaled so that their kinematic limits
match the kinematic limits for events in the $\omega$ signal region.
This procedure is necessary to avoid the introduction of artificial
structures in background-subtracted distributions for these
variables near the kinematic limits.

\begin{figure}[hbt]
\begin{center}
\epsfig{file=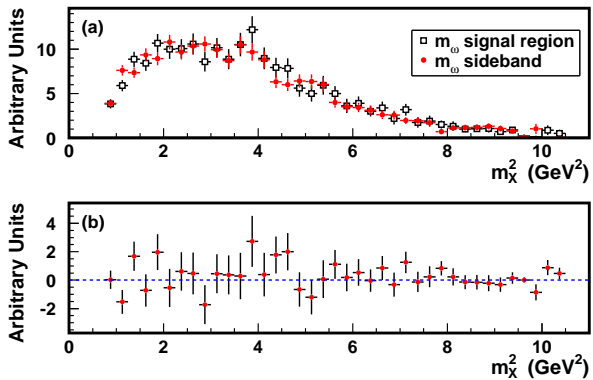,width=\linewidth}
\caption{\protect (a) Efficiency-corrected 
$m^2_X$ distributions for events from the $m_{ES}$ sideband 
with reconstructed $m_{\omega}$ in the signal
and sideband regions (arbitrary units).  The distribution
for events in the sideband region has been scaled by a factor
of $\frac{4}{7}$.
(b) Background subtracted $m^2_X$ distribution for events from the
$m_{ES}$ sideband (arbitrary units).  
This distribution has been obtained by subtracting 
the two distributions in (a).}
\label{fig:unbiased}
\end{center}
\end{figure}

We test the sideband subtraction algorithm on a number of background 
samples such as Monte Carlo $B\bar{B}$ events and data events in
sidebands of $\mes$ and $\Delta E$.  These tests are performed 
using the efficiency parametrization discussed in Section~\ref{sec:eff}.
We find that background-subtracted kinematic
distributions in the background samples show no significant structure.
One sample distribution is shown in Fig.~\ref{fig:unbiased}.
We find a small bias in the extraction
of the background-subtracted yields if the parameter $\beta$ in
Eq.~\ref{eq:bgsub-effcorr} is set to unity.  To correct for this 
bias we set $\beta = 0.975$, with an estimated systematic 
uncertainty of $\pm 0.010$.

\section{Efficiency Parametrization}
\label{sec:eff}

The process of interest ($\BtoDstOmgPi$) is
the three-body decay of a pseudoscalar particle into two vector particles 
and a pseudoscalar particle.  We parametrize the reconstruction efficiency 
as a function of five variables:

\begin{enumerate}

\item $d$, an index that labels 
the decay mode of the $D^0$; {\it i.e.}, $D^0 \to K^- \pi^+$,
$K^- \pi^+ \pi^+ \pi^-$, or $K^- \pi^+ \pi^0$;

\item $E_{\omega}$, the energy of the $\omega$ in the $\Bzb$ rest frame;

\item $E_{D^*}$, the energy of the $D^*$ in the $\Bzb$ rest frame;

\item $\cos\theta_D$, the cosine of the decay angle of the $D^*$; 
{\it i.e.}, the 
angle 
between 
the $D^0$ and the direction opposite the flight of the $\Bzb$ 
in the $D^{*+}$ rest frame.

\item $\cos\alpha$, the cosine of the angle
between the vector normal to the $\omega$ decay plane and the 
direction opposite the flight of the $\Bzb$, measured in the 
$\omega$ rest frame.

\end{enumerate}

Note that two other variables would be needed to fully describe the 
kinematics of the decay chain.  These are the angles that define, in 
addition to $\cos\theta_D$ and $\cos\alpha$,
the orientation between the decay 
planes of the $D^*$ and the $\omega$ and the decay plane of
the $\Bzb$.  Monte Carlo studies show that the reconstruction
efficiency is independent of these two variables.
The $E_{\omega}$ and $E_{D^*}$ variables are the usual Dalitz variables
used to describe three-body decays.  
Because of energy-momentum conservation the
$E_{\omega}$ and $E_{D^*}$ variables
are equivalent in information content to the squared invariant masses
of the $D^* \pi$ ($m^2_{D^*\pi}$) 
and $\omega \pi$ ($m^2_X$) systems respectively.

The efficiency is then parametrized as 
\begin{eqnarray}
\epsilon(\vec{x}_i) & = &
\epsilon(E_{\omega}, E_{D^*}, \cos\theta_D, |\cos\alpha|; d)\\ 
 & = & \epsilon'(E_{\omega}, E_{D^*}; d) \cdot
c_1(E_{\omega},|\cos\alpha|) \cdot
c_2(E_{D^*},\cos\theta_D; d). \nonumber
\end{eqnarray}

\begin{figure}[hbt]
\begin{center}
\mbox{\epsfig{figure=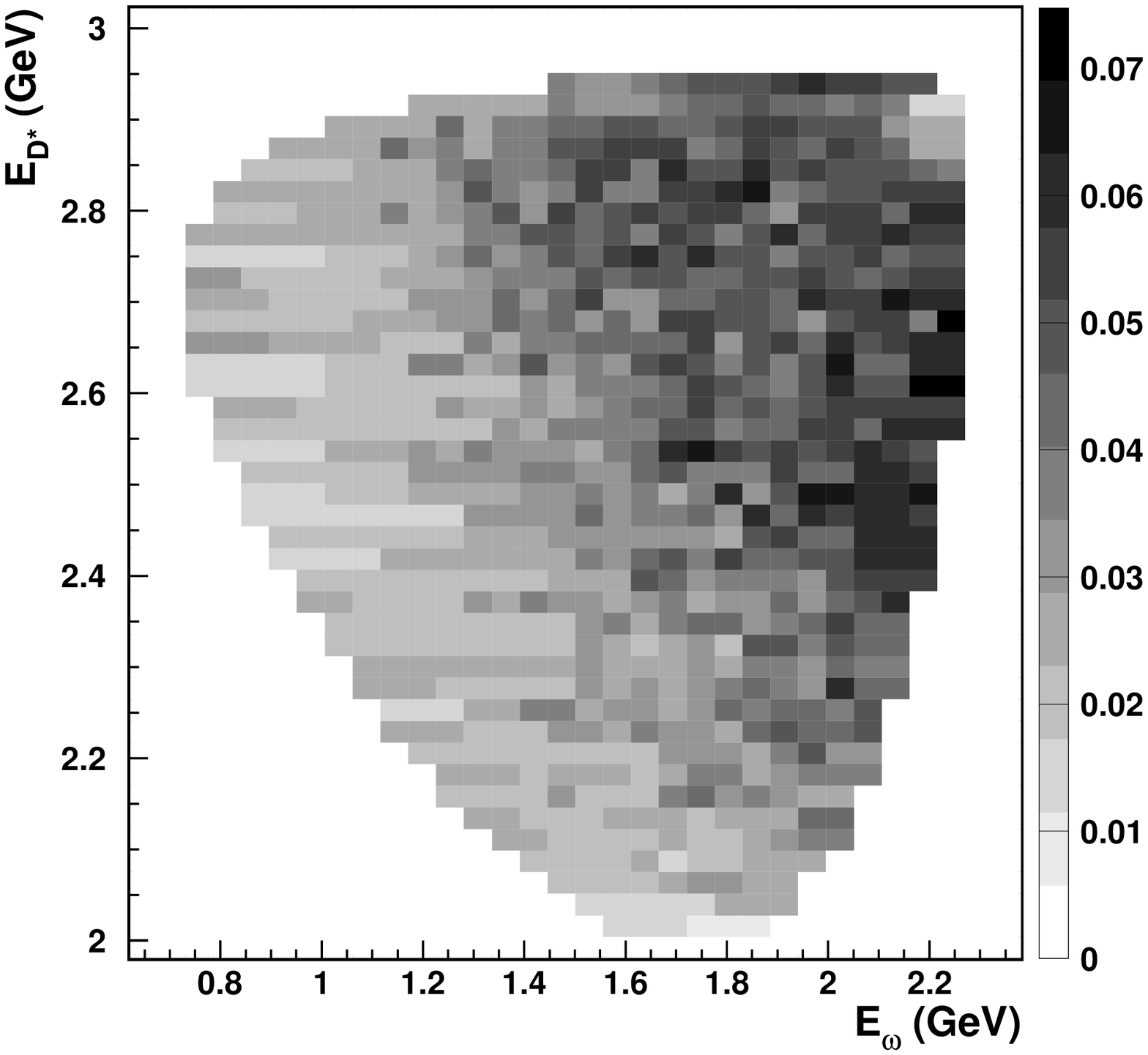,width=\linewidth}}
\caption{The $\epsilon'$($D^0 \to K^- \pi^+$, $E_{\omega}$, $E_{D^*}$) 
distribution for $\BtoDstOmgPi$ Monte Carlo events.} 
\label{fig:eff}
\end{center}
\end{figure}

\noindent The functions $\epsilon'$, $c_1$, and $c_2$ are extracted from
Monte Carlo simulations and tabulated as a set of 
two dimensional histograms. As an example, the $\epsilon'$ distribution 
for events with $D^0 \to K^- \pi^+$ is given in Fig.~\ref{fig:eff}.

The efficiency parametrization is validated using samples of Monte Carlo 
signal events.  These samples are generated with a variety of 
ad-hoc kinematic properties; {\it e.g.}, different polarizations
for the $D^*$ and the $\omega$, different shapes of the $m_X^2$ 
distribution.
In all cases we find that the 
shapes of kinematic distributions are well reproduced after 
the efficiency correction.

We use the following method to estimate the effect of the finite
statistics of the Monte Carlo sample.  We generate a set of
400 new $\epsilon'$, $c_1$, and $c_2$ templates based
on the nominal templates obtained from Monte Carlo signal events.
If the measured efficiency in a given bin of the nominal
template is $\mu \pm \sigma$, the corresponding efficiencies 
in the new templates are drawn from a Gaussian distribution 
of mean $\mu$ and standard deviation $\sigma$.
Then, the measurement of any quantity of interest ({\it e.g.}, $m_X^2$)
is repeated 400 times, according to Eq.~\ref{eq:bgsub-effcorr},
using the new templates.  The spread in the results obtained from 
events reconstructed in data is
a measure of the systematic uncertainty due to the finite number of 
available Monte Carlo events.  This spread is then 
added in quadrature to the statistical uncertainty of our results.

We observe a small bias in the total number of reconstructed 
signal events obtained from the efficiency correction.  This is due to the 
fact that, although the uncertainty on $\epsilon(\vec{x}_i)$ is Gaussian, 
the factor $1/\epsilon(\vec{x}_i)$ used in the 
efficiency correction procedure (Eqs.~\ref{eq:bgsub},~\ref{eq:bgsub-effcorr}) 
does not obey Gaussian statistics.  As a result, after applying the 
efficiency correction, the total number of reconstructed events 
tends to slightly overestimate the true value.

In order to quantify this bias on the nominal result due to the 
finite number of Monte Carlo signal events, we first determine the 
mean of the total number of reconstructed signal events in data for 
the 400 new efficiency templates.  This mean differs from the nominal 
result by a few percent ($\delta$).  
We then repeat the 
procedure described above using events reconstructed from signal Monte Carlo.
We use the results of these Monte Carlo studies to describe the bias as a 
function of $\delta$.
We find that after applying the efficiency correction and subtracting the 
$m_{\omega}$ sideband, the total number of events reconstructed using signal 
Monte Carlo exceeds the true value by $(0.6 \pm 0.4)\cdot \delta$.  
We correct our final results by this amount.

\section{Results}
\label{sec:results}

We use the procedure outlined above, with one additional correction, 
to extract the branching fraction, the
$m_X^2$ distribution,
the Dalitz plot distribution, the $m_{D^* \pi}$ distribution, 
and the polarization of the $D^*$ as a function of $m_X$.
The one additional correction accounts for the region of phase space 
with low acceptance that was excluded from the analysis.  This region
corresponds to values of $\cos\theta_D$ near 1 for low $E_D^*$, 
or equivalently high $m_X^2$.  This correction factor
varies between approximately 1.2 at $m_X^2 = 8$ GeV$^2$ 
and 1.6 at $m_X^2 = 11$ GeV$^2$.  Since most of the data is
at $m_X^2 < 4$ GeV$^2$, the combined effect of this correction is 
quite small; it amounts to an increase of less than 1\% relative to 
the measured branching fraction.

For the branching fraction, we find
${\cal B}(\BtoDstOmgPi) = 
(2.88 \pm 0.21 {\rm (stat.)} \pm 0.31 {\rm(syst.)}) \times 10^{-3}$.
The total systematic uncertainty of 10.8\% arises from the following sources:

\begin{itemize}

\item The uncertainties in the branching fractions of the $D^*$, $D$, 
and $\omega$: 5\%.

\item The uncertainty in the reconstruction efficiency of neutral
pions at \babar, which is estimated to be 3\% per $\pi^0$.
This amounts to a 6\% uncertainty for events reconstructed with 
$D^0 \to K^- \pi^+ \pi^0$, and 3\% for the other modes.  Combining 
these modes, the systematic uncertainty from this source is 4.3\%.

\item The uncertainty in the reconstruction efficiency for charged 
tracks.  From a variety of control samples, this is estimated to
be 0.6\% (0.8\%) for each track of transverse momentum above (below) 
200 MeV.  Including the uncertainty for the 
low momentum pion produced in the $\Dstarp$ decay, we obtain 
a systematic uncertainty of 5.3\%.

\item The uncertainty in the efficiency of the kaon particle identification
requirements: 2\%.

\item The uncertainty due to the limited Monte Carlo sample size in the
efficiency calculation: 3.8\%.

\item The uncertainty on the quantity $\beta$ in 
Eq.~\ref{eq:bgsub-effcorr}: 2.6\%.

\item The uncertainty in the efficiency of the various event selection
criteria, estimated to be 4.3\%.

\item The uncertainty in the number of $B\bar{B}$ events in the 
\babar\ event sample: 1.1\%.

\item The uncertainty in the correction due to the removal of 
events at high $\cos\theta_{D}$ and small $E_{D^*}$: 0.3\%.

\end{itemize}

Some of these systematic uncertainties vary as a function of $m_X^2$.  
For example, the uncertainty on the correction due to removing a region of 
$(E_{D^*},\cos \theta_D)$ phase space is only relevant to 
events with $m_X^2$ above 8 GeV$^2$.  A portion of the systematic uncertainty 
due to limited Monte Carlo sample size also varies as a function of $m_X^2$.
Therefore, quantities measured as a function of $m_X^2$
include a common scale uncertainty of 10.5\%
and a systematic uncertainty that varies with $m_X^2$ and is 
typically below a few percent.

\begin{figure}[hbt]
\begin{center}
\mbox{\epsfig{figure=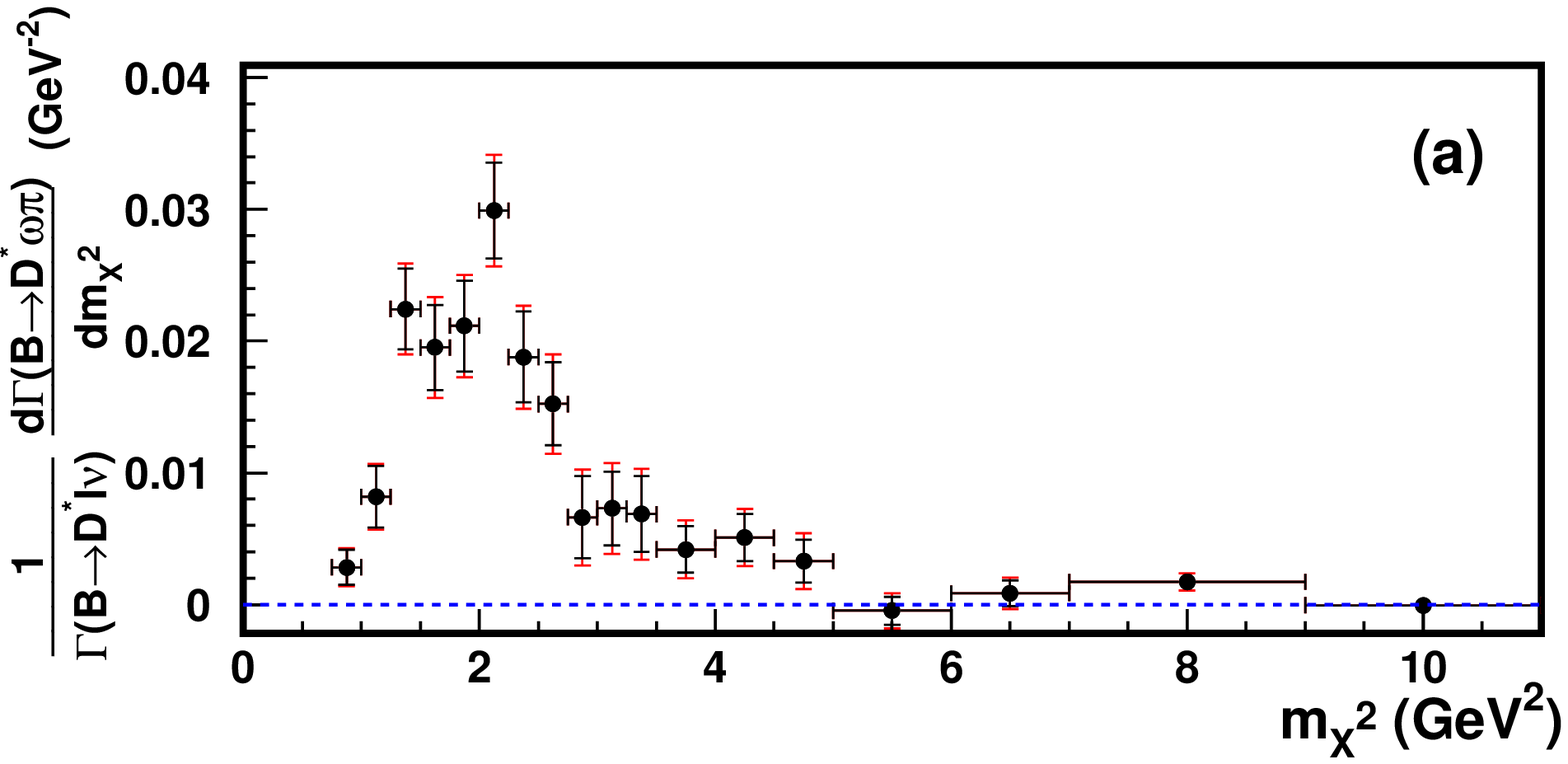,width=\linewidth}}
\mbox{\epsfig{figure=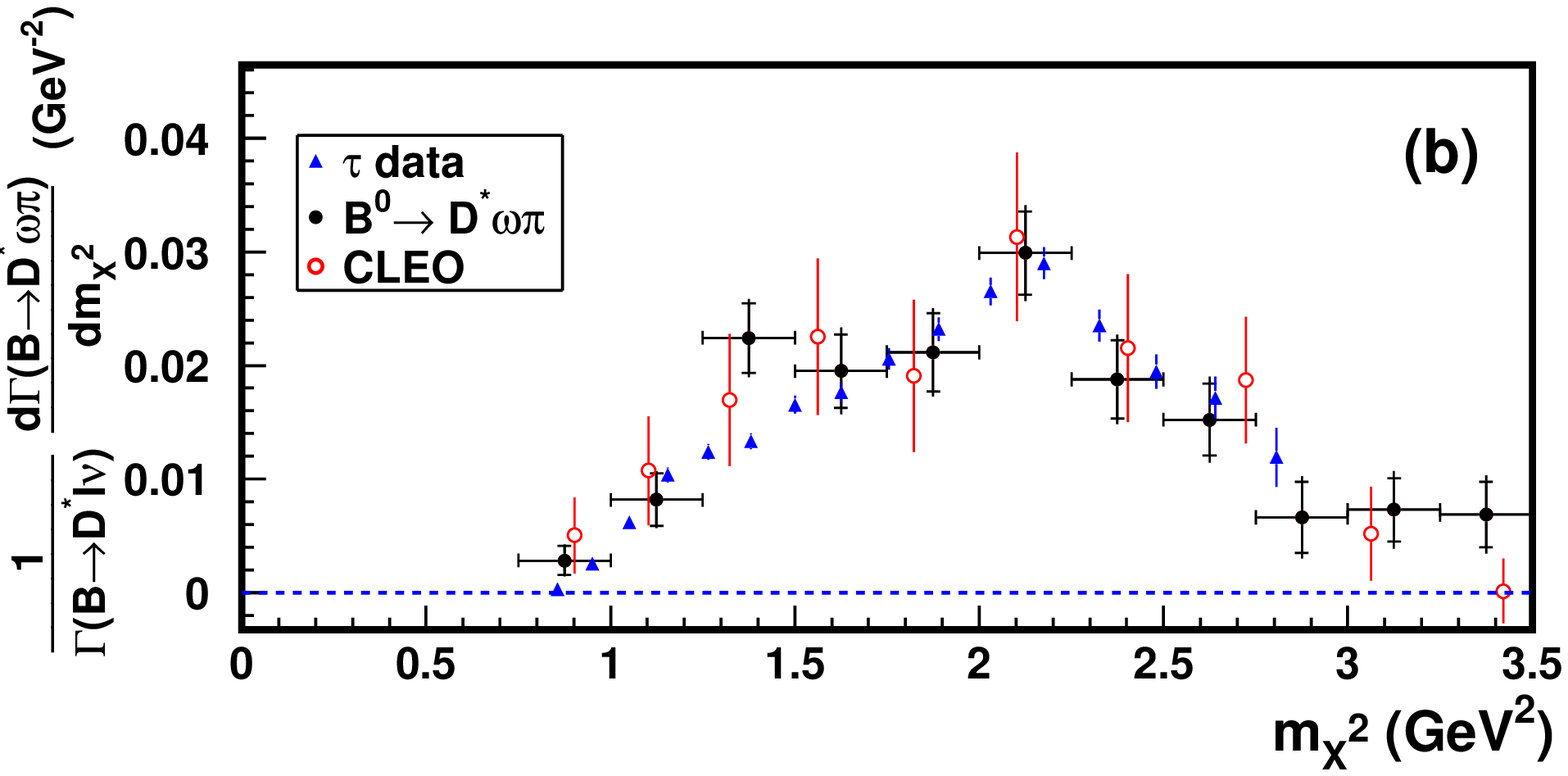,width=\linewidth}}
\caption{\protect (a) Data $m_X^2$ distribution normalized to the 
semileptonic width $\Gamma(B \to D^* \ell \nu)$.  The inner
error bars reflect the statistical uncertainties on the data.  The total 
error bars include the $m_X^2$-dependent systematic uncertainties.
A common 11.3\% scale systematic uncertainty is not shown.
(b) Same as (a) but zoomed-in on the low $m_X^2$ region, where 
comparisons based on factorization and $\tau$ data can be made. 
Also shown here are the results from the CLEO analysis~\cite{CLEO}.}
\label{fig:mx2}
\end{center}
\end{figure}

The $m^2_X$ distribution, normalized to the semileptonic width
$\Gamma(B \to D^* \ell \nu)$~\cite{PDG},
is shown in Fig.~\ref{fig:mx2}.  A scale uncertainty on our result of 11.3\% 
is not shown.  This uncertainty combines 
a 4.2\% uncertainty in $\Gamma(B \to D^* \ell \nu)$ with the 10.5\% 
uncertainty from the sources listed above.
The bulk of the data is concentrated in a broad peak around 
$m_X^2 \approx 2$ GeV$^2$, in the region of 
$\rho(1450) \to \omega \pi$.
Our distribution agrees 
well in both shape and normalization
with predictions based on factorization and $\tau$ decay 
data~\cite{CLEOtau}
in the region $m_X^2 \leq 2.8$~GeV$^2$ covered by the tau
data.

\begin{figure}[bth]
\begin{center}
\epsfig{figure=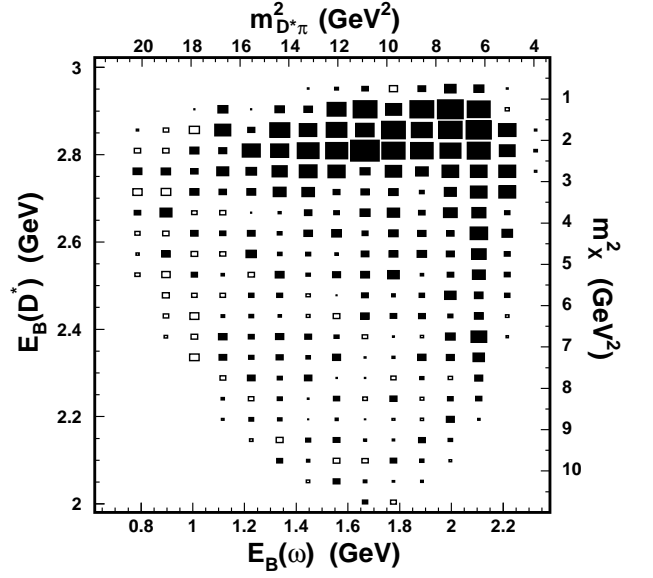,width=\linewidth}
\caption{\protect Background-subtracted and efficiency-corrected
Dalitz plot for $B \to D^* \omega \pi$.  The relative box sizes
indicate the population of the bins.  Black boxes indicate positive
values, white boxes indicate negative values, which can occur because
of statistical fluctuations in the subtraction procedure.}
\label{fig:dalitz}
\end{center}
\end{figure}

The background-subtracted and efficiency-corrected Dalitz plot 
is shown in Fig.~\ref{fig:dalitz}.  
One notable feature of the decay distribution
is an enhancement for $D^*\pi$ masses near 2.5~GeV
($m^2_{D^*\pi} \sim 6.3$ GeV$^2$).
The enhancement occurs in the region where one expects
to find a broad $J=1$ $D^{**}$ resonance ($D'_1$)
that decays
via S-wave to $D^* \pi$.  Thus, this enhancement could
be due to the color-suppressed decay
$\Bzb \to D'_1 \omega$, followed by
$D'_1 \to D^{*+} \pi^-$.

\begin{figure}[bth]
\begin{center}
\epsfig{figure=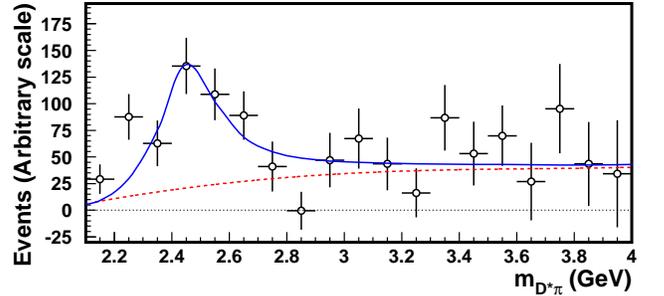,width=\linewidth}
\caption{\protect Background-subtracted and 
efficiency-corrected $D^*\pi$ mass distribution with
$\cos\theta_{D^*} < 0.5$.  The superimposed fit is 
described in the text.}
\label{fig:ddstar}
\end{center}
\end{figure}

In Fig.~\ref{fig:ddstar} we show the 
background-subtracted and efficiency-corrected $D^*\pi$ mass distribution
for events away from the $\rho(1450)$ peak, fitted to the sum
of a fourth order polynomial and a relativistic Breit-Wigner.
In this figure, in order to remove the contribution from the $\rho(1450)$, 
we have required $\cos\theta_{D^*} < 0.5$, where 
$\theta_{D^*}$ is the angle between the momentum 
of the $D^*$ in the $D^*\pi$ rest frame, and the flight direction of
$D^*\pi$ system.  
We use the $\cos \theta_{D^*}$ variable rather than $m_X^2$
to remove the $\rho(1450)$ contribution because 
the distribution in $\cos \theta_{D^*}$ is uniform
for S-wave $D'_1 \to D^* \pi$ decay.  The yield of possible 
$\Bzb \to  D'_1 \omega$ events in
Fig.~\ref{fig:ddstar} can then be easily extrapolated to the full kinematic
range.  Furthermore, by subdividing the dataset in bins of
$\cos \theta_{D^*}$ we can test the S-wave decay hypothesis.

The fitted mass
and width of the Breit-Wigner in Fig.~\ref{fig:ddstar} 
are $2477 \pm 28$ MeV and
$266 \pm 97$ MeV, respectively.  These values are 
consistent with the parameters of the broad $D'_1 \to D^* \pi$ 
resonance observed by the Belle collaboration in
$B \to D'_1 \pi$ decays, $m = 2427 \pm 36$ MeV
and $\Gamma = 384^{+107}_{-75} \pm 74$ MeV~\cite{Belle}.
We have also split the data set of Fig.~\ref{fig:ddstar}
into three equal-sized bins of $\cos\theta_{D^*}$.
We find that the fitted amplitude of the Breit-Wigner
component is the same, within statistical uncertainties, in the three 
data sets.  This is consistent with expectations 
for an S-wave $D'_1 \to D^* \pi$ decay.

If we assume that the enhancement for $D^*\pi$ masses near 2.5 GeV
is actually due to $\Bzb \to D'_1 \omega$, $D'_1 \to D^{*+} \pi^-$,
we extract the branching fraction

\begin{eqnarray}
{\cal B}(\Bzb \to D'_1 \omega) \times 
{\cal B}(D'_1 \to D^{*+} \pi^-) = \nonumber \\
(4.1 \pm 1.2 \pm 0.4 \pm 1.0) \times 10^{-4}. 
\label{eq:br_dds}
\end{eqnarray}

\noindent In this measurement, the first uncertainty is statistical, 
the second uncertainty is from the uncertainties in common with the 
${\cal B}(\BtoDstOmgPi)$ measurement,
and the last uncertainty arises from the uncertainties on the choice of 
the nonresonant shape in Fig.~\ref{fig:ddstar} (10\%) and the 
uncertainties in the parameters of the $D'_1$ resonance (22\%).
This branching fraction has been obtained from fitting the sample
of events with $\cos\theta_{D^*} < 0.5$, and scaling up the result
by a factor of $\frac{4}{3}$.  This procedure neglects interference
effects between $\Bzb \to D'_1 \omega$ and $\Bzb \to D^* \omega \pi$.

The branching fraction in Eq.~\ref{eq:br_dds} is comparable
to the branching fractions for $\Bzb \to D^{(*)0} \omega$~\cite{PDG}.
Also, we see no evidence for decays into the two narrow $D^{**}$
resonances at 2420 and 2460 MeV.  This is in contrast to the
color-favored $\B^- \to D^{**0} \pi^-$ decays, 
where the three $D^{**}$ modes contribute with comparable strengths,
and where the $B^- \to D'_1 \pi^-$ branching fraction
is one order of magnitude
smaller than that of $B^- \to D^{(*)0} \pi^-$.

The presence of $\Bzb \to D'_1 \omega$ would affect the 
comparison of the data with the theoretical predictions of 
Fig.~\ref{fig:mx2}.  As can be seen in Fig.~\ref{fig:dalitz},
$\Bzb \to D'_1 \omega$ would mostly contribute at high value of $m_X^2$,
while the factorization test can be carried out only where the
$\tau$ data is available; {\it i.e.}, for $m_X^2 < 3$ GeV$^2$.
Based on the estimated branching fraction of $\Bzb \to D'_1 \omega$,
and neglecting interference effects, the contribution
of $\Bzb \to D'_1 \omega$ to the $m_X^2 < 3$ GeV$^2$ distribution 
would be less than 5\%.

If the decay $\BtoDstOmgPi$ proceeds dominantly
through $\Bzb \to D^{*+} \rho(1450)$, $\rho(1450) \to \omega \pi^-$,
a measurement of the polarization of the $D^*$ can provide a further
test of factorization and HQET~\cite{poltheory}.
The angular distribution in the $D^{*+} \to D^0 \pi^+$ decay
can be written as a function of
three complex
amplitudes $H_0$ (longitudinal), and $H_+$ and
$H_-$(transverse), as

\begin{equation}
\frac{d\Gamma}{d\cos\theta_{D}} \propto 
4|H_0|^2 \cos^2\theta_{D} + (|H_{+}|^2 + |H_{-}|^2)\sin^2\theta_{D},
\label{eq:dist}
\end{equation}

\noindent where $\theta_{D}$ is the decay angle of the $D^*$ defined above. 

The longitudinal polarization fraction $\Gamma_L/\Gamma$, given by
\begin{equation}
\frac{\Gamma_L}{\Gamma} = \frac{|H_0|^2}{|H_0|^2+|H_{+}|^2 + |H_{-}|^2}, 
\label{eq:gamma}
\end{equation}

\noindent can then be extracted using Eq.~\ref{eq:dist}
from a fit to the angular distribution in the decay of the $D^*$.

We divide our dataset in ranges of $m^2_X$, and perform binned chi-squared
fits to the efficiency-corrected, background-subtracted, 
$D^*$-decay angular distributions.  In these measurements, nearly all 
of the systematic uncertainties discussed above cancel.  
As a result, the $m_X^2$-dependent uncertainty due to the finite 
Monte Carlo sample is the dominant systematic uncertainty, 
and typically results in an uncertainty on 
$\Gamma_L/\Gamma$ at the few percent level.  We also include 
a systematic uncertainty due to the parameter $\beta$ in 
Eq.~\ref{eq:bgsub-effcorr}.  This uncertainty is about one order
of magnitude smaller.

\begin{figure}[htb]
\begin{center}
\mbox{\epsfig{figure=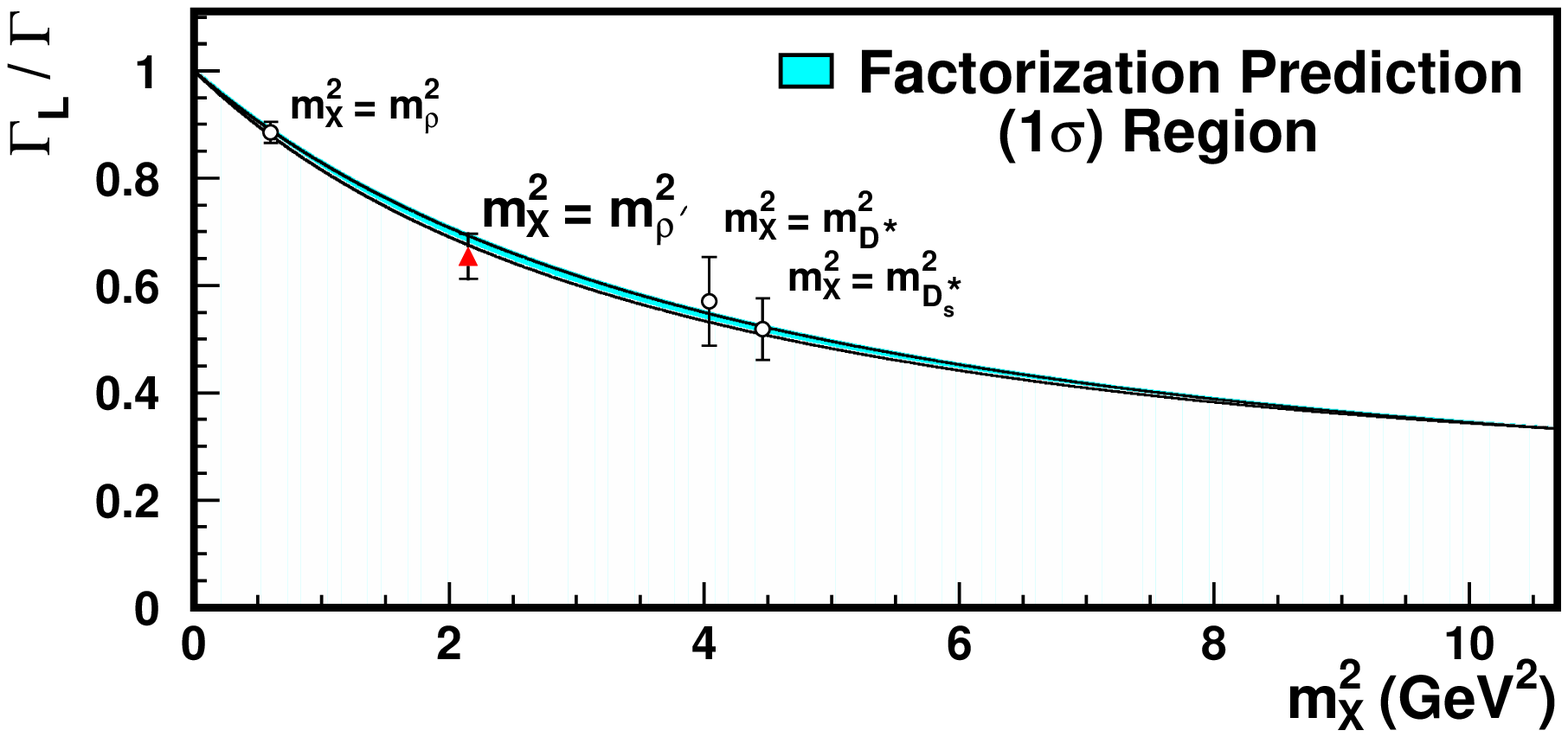,width=\linewidth}}
\end{center}
\caption{The fraction of longitudinal polarization as a function of $m_X^2$, 
where $X$ is a vector meson.  Shown (as a triangle) is the $\BtoDstOmgPi$ 
polarization measurement for events with $1.1 < m_X < 1.9$ GeV 
($m_X^2$ = $m^2_{\rho'}$, where $\rho' \equiv \rho(1450)$), 
as well as earlier measurements (indicated by open circles) of 
$\Bzb \to D^{*+} \rho^-$~\cite{DstRho}, $\Bzb \to D^{*+} D^{*-}$~\cite{DstDst}, 
and $\Bzb \to D^{*+} D_s^{*-}$~\cite{DstDSst}.
The shaded region represents the prediction ($\pm$ one standard deviation) 
based on factorization and HQET, extrapolated from the semileptonic 
$\Bzb \to D^{*+} \ell^- \bar{\nu}$ form factor results~\cite{DstlnuFF}.}
\label{fig:polfig}
\end{figure}

The measured longitudinal
polarization fractions as a function of $m_X$ are 
presented in Table I.  
Near the mean of the $\rho(1450)$ resonance ($1.1 < m_X < 1.9$ GeV), we
find $\Gamma_L/\Gamma = 0.654 \pm 0.042 {\rm(stat.)} \pm 
0.016 {\rm(syst.)}$.
This result is in agreement with the 
previous result in the same mass range from the CLEO collaboration,
$\Gamma_L/\Gamma = 0.63 \pm 0.09$.  It is also in agreement
with predictions based on HQET, factorization, and the 
measurement  of semileptonic $B$-decay form factors,
$\Gamma_L/\Gamma = 0.684 \pm 0.009$~\cite{DstlnuFF}, assuming that the decay 
proceeds via $\Bzb \to D^{*+} \rho(1450)$, $\rho(1450) \to \omega \pi^-$.  These 
results are shown in Fig.~\ref{fig:polfig}.

\begin{table}[h]
\label{tab:Dst-polTable}
\begin{center}
\caption{Results of the $D^*$ polarization measurement in bins of $m_X$.  
The first uncertainty is statistical and the second is systematic.}
\begin{tabular}{lc} \hline
$m_X$ range (GeV) &  $\Gamma_L / \Gamma$ \\ \hline
below 1.1         &  0.458 $\pm$ 0.189 $\pm$ 0.059  \\ 
1.1 - 1.35        &  0.779 $\pm$ 0.062 $\pm$ 0.020  \\ 
1.35 - 1.55       &  0.733 $\pm$ 0.071 $\pm$ 0.024  \\ 
1.55 - 1.9        &  0.435 $\pm$ 0.102 $\pm$ 0.040  \\
1.9 - 2.83        &  0.656 $\pm$ 0.182 $\pm$ 0.077  \\ \hline
\end{tabular}
\end{center}
\end{table}

\section{Conclusions}

We have studied the decay $\BtoDstOmgPi$ with 
a larger data sample than previously available.
We measure the branching fraction 
to be ${\cal B}(\BtoDstOmgPi) =
(2.88 \pm 0.21$(stat.) $\pm 0.32$(syst.)$) \times 10^{-3}$.  

The invariant mass spectrum of the $\omega \pi$ system is 
found to be in agreement with theoretical expectations based
on factorization and $\tau$ decay data.  The Dalitz plot 
for this mode is very non-uniform, with most of the rate
at low $\omega \pi$ mass.  
We also find an enhancement
for $D^* \pi$ masses broadly distributed around 2.5 GeV.
This enhancement could be due to color-suppressed 
decays into the broad $D'_1$ resonance, 
$\Bzb \to D'_1 \omega$, followed by $D'_1 \to D^{*+} \pi^-$,
with a branching fraction comparable to 
$\Bzb \to D^{(*)0} \omega$.

We also measure the fraction of $D^*$ longitudinal polarization 
in this decay.  In the region of $\omega \pi$ mass between
1.1 and 1.9 GeV, where one expects contributions from
$\Bzb \to D^{*+} \rho(1450)$, $\rho(1450) \to \omega \pi^-$,
we find 
$\Gamma_L/\Gamma = 0.654 \pm 0.042 {\rm(stat.)} \pm 
0.016 {\rm(syst.)}$, 
in agreement
with predictions based on HQET, factorization, and the 
measurement  of semileptonic $B$-decay form factors.

\smallskip

We are grateful for the 
extraordinary contributions of our \pep2\ colleagues in
achieving the excellent luminosity and machine conditions
that have made this work possible.
The success of this project also relies critically on the 
expertise and dedication of the computing organizations that 
support \babar.
The collaborating institutions wish to thank 
SLAC for its support and the kind hospitality extended to them. 
This work is supported by the
US Department of Energy
and National Science Foundation, the
Natural Sciences and Engineering Research Council (Canada),
Institute of High Energy Physics (China), the
Commissariat \`a l'Energie Atomique and
Institut National de Physique Nucl\'eaire et de Physique des Particules
(France), the
Bundesministerium f\"ur Bildung und Forschung and
Deutsche Forschungsgemeinschaft
(Germany), the
Istituto Nazionale di Fisica Nucleare (Italy),
the Foundation for Fundamental Research on Matter (The Netherlands),
the Research Council of Norway, the
Ministry of Science and Technology of the Russian Federation, and the
Particle Physics and Astronomy Research Council (United Kingdom). 
Individuals have received support from 
CONACyT (Mexico), the Marie-Curie Intra European Fellowship program (European Union),
the A. P. Sloan Foundation, 
the Research Corporation,
and the Alexander von Humboldt Foundation.

\end{document}